\documentclass[review,12pt,authoryear]{elsarticle}
\date{}
\usepackage{subfigure}
\usepackage{amssymb}

\begin{document}
\begin{frontmatter}

\title{Comment on \emph{Increased mortality is predicted of Inachis io larvae caused by Bt-maize
pollen in European farmland},  Ecological Modelling, 2013}
\author[FC] {Francesco Camastra}
\ead{camastra@ieee.org}
\author[FC]{Angelo Ciaramella}
\ead{angelo.ciaramella@ieee.org}
\author[FC]{Antonino Staiano\corref{cor1}} 
\ead{staiano@ieee.org}
\cortext[cor1]{Corresponding author}
\address[FC]{Department of Science and Technology, University of Naples Parthenope, \\ Centro Direzionale Isola C4, 80143 Naples, Italy}

\begin{abstract}

In this comment, we  analyze  the  mathematical model by Holst et al that estimates  the increased mortality of \emph{Inachis io} larvae caused by Bt-maize pollen in European farmaland.
In particular, we focus our attention on the part of Holst et al's work that models the Bt-maize pollen
loss on plant leaves by a differential equation. We provide the exact solution of the differential equation, showing that the behaviour of the solution, when the time goes to infinity, is different from what claimed in Holst et al's work.   

\end{abstract}
\end{frontmatter}
\section{Introduction}
Nowadays, the cultivation of Genetically Modified Plants (GMP) is very
widespread in the world, in particular in America and Asia. On the other side, the debate in the scientific community and in the public opinion about the GMP effects is becoming harder and harder \citep{Sanvido11}.
To this purpose, recently appeared some mathematical models for estimating the effects of Bt-maize \citep{Perry10, Perry12, Holst13}, on non-target Lepidoptera. In this comment, we  analyze  the  mathematical model by Holst et al \citep{Holst13} that estimates the increased mortality of \emph{Inachis io} larvae caused by Bt-maize pollen in farmaland of Northern Europe.
In particular, we focus our attention on the part of Holst et al's work that models the Bt-maize pollen
exposure of Inachis io larvae. The work is organized as follows: 
In Section \ref{Bt-maize exposure} the Bt-maize pollen
exposure of Inachis io larvae model is discussed; 
finally, some conclusions are drawn in Section \ref{concl}. 
\section{Bt-maize Pollen exposure of Inachis io larvae}\label{Bt-maize exposure}
In their model, Holst et al represents the pollen exposure in terms of pollen concentration in
the environment. In the specific case of Inachis io larvae pollen exposure is expressed in 
terms of average pollen density, measured in cm$^{-2}$, on the leaves of the food plant. Pollen deposition rate follows a parabolic curve through the pollination period \citep{Kawashima04}.  The integral of the curve provides 
the accumulated deposition. However, pollen deposited on food plant leaves will be lost after deposition. Holst et al model 
pollen loss on plant leaves  by means of the following
differential equation:
\begin{equation}\label{eq0}
\frac{dN}{dt}= a t^2+b t+c - \epsilon N,
\end{equation}
where $N$ is pollen density (expressed  in cm$^{-2}$),  $t$ is time (measured in days), $\epsilon$ is \emph{loss rate}, i.e., relative amount of pollen lost per day, and $a, b, c$ are the coefficients of the parabola.
\\
Holst et al declare that \emph{the equation (\ref{eq0}) was integrated and re-parameterized to predict 
average pollen density on leaves ($N$)} and show, in  figure 2 of the manuscript, that the average pollen density have a maximum at $N_{peak}$ and goes to zero when the time $t \rightarrow \infty$.
\\
Having said that, we observe that the equation (\ref{eq0}) is a 
\emph{inhomogeneous linear differential equation of  first order} \citep{Korn00} and can be 
solved exactly.
To this purpose, we recall that a inhomogeneous linear differential equation of  first order has 
the following  form:
\begin{equation}\label{eqa}
\frac{dN}{dt}=  u(t) N + v(t),
\end{equation}
and has as general solution:
\begin{equation}
N(t)= e^{\int_{\alpha}^t u(x)dx}[K+\int_{\beta}^t v(x) e^{-\int_{\alpha}^x u(s)ds} dx],
\end{equation}
where $K \in \mathbb{R}$ and $\alpha, \beta$ are values that have to be chosen in a properly 
way that for convenience in our case we put  to zero.
Since in the case of equation (\ref{eq0}) it is  
\begin{equation}\label{eqb}
u(t)=-\epsilon, \hspace{2cm} v(t)=a t^2+b t+c;
\end{equation} 
the solution of the differential equation (\ref{eq0}) is:
\begin{eqnarray}
N(t) & = &  e^{-\epsilon t} [K+\int_{0}^t (a x^2+b x+c) e^{\epsilon x} dx]  \nonumber \\
        & = & K e^{-\epsilon t}+ e^{-\epsilon t}[e^{\epsilon x}\frac{a (2-2x\epsilon+x^2\epsilon^2) +
\epsilon(c\epsilon+b(-1+x\epsilon))}{\epsilon^3}]_0^t \nonumber \\
        & = & \frac{a}{\epsilon}t^2+(-\frac{2a}{\epsilon^2}+\frac{b}{\epsilon})t+(\frac{2a}{\epsilon^3}+\frac{c}{\epsilon}-\frac{b}{\epsilon^2})+ (K-\frac{2a-b\epsilon+c \epsilon^2}{\epsilon^3})e^{-\epsilon t}
\end{eqnarray}
where $K \in \mathbb{R}$.
\\
It is immediate to verify that  the  limit of pollen density $N(t)$, when $t \rightarrow \infty$  is not  0,
as shown in figure 2 of Holst et al's manuscript, but:    
\begin{equation}
\displaystyle\lim_{t \rightarrow \infty} N(t) = \left\{
\begin{array}{ll}
\frac{c}{\epsilon} & \mbox{if $a=b=0$}  \\
\infty & \mbox{otherwise}  
\end{array}\right\}
\end{equation}
The degenerate case    $a=b=0$  has no practical interest,  since the parabola $v(t)=at^2+bt+c$ reduces to a  constant line $v(t)=c$. In all other cases, pollen density $N(t)$ goes to infinity
when time $t \rightarrow \infty$. Since it is clearly non acceptable, this means that the differential equation (\ref{eq0}) cannot model correctly pollen density on the food plant. Therefore the theoretical soundness of Holst et al' s model is mined remarkably,  in our opinion.
\section{Conclusion}\label{concl}
In this paper, we  have discussed  the  mathematical model by Holst et al that estimates  the increased mortality of \emph{Inachis io} larvae caused by Bt-maize pollen in farmaland of Northern Europe. We have paid particular attention on the differential equation that  Holst et al 
use to  model the Bt-maize 
pollen loss,  providing its exact solution.  We showed that the behaviour of the solution, when the time goes to infinity, is different from what claimed in Holst et al's work. 
Since the differential equation does not model correctly the pollen density on the food plant,  
 the theoretical soundness of Holst et al' s work seems to be mined remarkably.
\bibliographystyle{elsarticle-num-names}
\bibliography{comment_holst}

\begin{thebibliography}{6}
\providecommand{\natexlab}[1]{#1}
\providecommand{\url}[1]{\texttt{#1}}
\providecommand{\urlprefix}{URL }
\expandafter\ifx\csname urlstyle\endcsname\relax
  \providecommand{\doi}[1]{doi:\discretionary{}{}{}#1}\else
  \providecommand{\doi}[1]{doi:\discretionary{}{}{}\begingroup
  \urlstyle{rm}\url{#1}\endgroup}\fi
\providecommand{\bibinfo}[2]{#2}

\bibitem[{Sanvido et~al.(2011)Sanvido, Romeis, and Bigler}]{Sanvido11}
\bibinfo{author}{O.~Sanvido}, \bibinfo{author}{J.~Romeis},
  \bibinfo{author}{F.~Bigler}, \bibinfo{title}{Environmental change challanges
  decision-making during post-market environmental monitoring of transgenic
  crops}, \bibinfo{journal}{Transgenic Research} \bibinfo{volume}{20}
  (\bibinfo{year}{2011}) \bibinfo{pages}{1191--1201}.

\bibitem[{Perry et~al.(2010)Perry, Devos, Arpaia, Bartsch, Gathmann, Hails,
  Kiss, Lheureux, Manachini, Mestdagh, Neemann, Ortego, Schiemann, and
  Sweet}]{Perry10}
\bibinfo{author}{J.~N. Perry}, \bibinfo{author}{Y.~Devos},
  \bibinfo{author}{S.~Arpaia}, \bibinfo{author}{D.~Bartsch},
  \bibinfo{author}{A.~Gathmann}, \bibinfo{author}{R.~S. Hails},
  \bibinfo{author}{J.~Kiss}, \bibinfo{author}{K.~Lheureux},
  \bibinfo{author}{B.~Manachini}, \bibinfo{author}{S.~Mestdagh},
  \bibinfo{author}{G.~Neemann}, \bibinfo{author}{F.~Ortego},
  \bibinfo{author}{J.~Schiemann}, \bibinfo{author}{J.~B. Sweet},
  \bibinfo{title}{A mathematical model of exposure of non-target Lepidoptera to
  Bt-maize pollen expressing Cry1Ab within Europe},
  \bibinfo{journal}{Proceedings of the Royal Society B, Biological Sciences}
  \bibinfo{volume}{277} (\bibinfo{year}{2010}) \bibinfo{pages}{1417--1425}.

\bibitem[{Perry et~al.(2012)Perry, Devos, Arpaia, Bartsch, Ehlert, Gathmann,
  Hails, Hendriksen, Kiss, Messean, Mestdagh, Neemann, Nuti, Sweet, and
  Tebbe}]{Perry12}
\bibinfo{author}{J.~N. Perry}, \bibinfo{author}{Y.~Devos},
  \bibinfo{author}{S.~Arpaia}, \bibinfo{author}{D.~Bartsch},
  \bibinfo{author}{C.~Ehlert}, \bibinfo{author}{A.~Gathmann},
  \bibinfo{author}{R.~S. Hails}, \bibinfo{author}{N.~B. Hendriksen},
  \bibinfo{author}{J.~Kiss}, \bibinfo{author}{A.~Messean},
  \bibinfo{author}{S.~Mestdagh}, \bibinfo{author}{G.~Neemann},
  \bibinfo{author}{M.~Nuti}, \bibinfo{author}{J.~B. Sweet},
  \bibinfo{author}{C.~C. Tebbe}, \bibinfo{title}{Estimating the effects of
  Cry1F Bt-maize pollen on non-target Lepidoptera using a mathematical model of
  exposure}, \bibinfo{journal}{Journal of Applied Ecology} \bibinfo{volume}{49}
  (\bibinfo{year}{2012}) \bibinfo{pages}{29--37}.

\bibitem[{Holst et~al.(2013)Holst, Lang, Lovei, and Otto}]{Holst13}
\bibinfo{author}{N.~Holst}, \bibinfo{author}{A.~Lang},
  \bibinfo{author}{G.~Lovei}, \bibinfo{author}{M.~Otto},
  \bibinfo{title}{Increased mortality is predicted of Inachis io larvae caused
  by Bt-maize pollen in European farmland}, \bibinfo{journal}{Ecological
  Modelling} \bibinfo{volume}{250} (\bibinfo{year}{2013})
  \bibinfo{pages}{126--133}.

\bibitem[{Kawashima et~al.(2004)Kawashima, Matsuo, Du, Takahashi, Inoue, and
  Yonemura}]{Kawashima04}
\bibinfo{author}{S.~Kawashima}, \bibinfo{author}{K.~Matsuo},
  \bibinfo{author}{M.~Du}, \bibinfo{author}{Y.~Takahashi},
  \bibinfo{author}{S.~Inoue}, \bibinfo{author}{S.~Yonemura}, \bibinfo{title}{An
  algorithm for estimating potential deposition of corn pollen for
  environmental assessment}, \bibinfo{journal}{Environmental Biosafety
  Research} \bibinfo{volume}{3} (\bibinfo{year}{2004})
  \bibinfo{pages}{197--207}.

\bibitem[{Korn and Korn(2000)}]{Korn00}
\bibinfo{author}{G.~A. Korn}, \bibinfo{author}{T.~A. Korn},
  \bibinfo{title}{Mathematical Handbook for Scientists and Engineers},
  \bibinfo{publisher}{Dover Publications}, \bibinfo{year}{2000}.

\end{thebibliography}
%
\end{document}